\documentclass[debug,overfull]{epl}

\usepackage{graphics}
\usepackage{graphicx}
\usepackage{pstricks}
\usepackage{epsfig}

\title{Quark masses without Yukawa hierarchies}

\author{H. Fanchiotti\inst{1}, C. Garc\'\i a-Canal\inst{1} and W. A. 
Ponce\inst{2}\thanks{E-mail: \email{wponce@naima.udea.edu.co}}}
\shortauthor{H. Fanchiotti \etal}
\institute{\inst{1}Laboratorio de F\'\i sica Te\'orica, Departamento de
F\'\i sica, Facultad de Ciencias Exactas, Universidad Nacional de La Plata, C.C. 67-1900, La
Plata Argentina.\\
\inst{2}Instituto de F\'\i sica, Universidad de Antioquia, A.A. 1226, Medell\'\i n, Colombia.}


\begin{document}

\maketitle

\begin{abstract}
{A model based on the local gauge group $SU(3)_c\otimes SU(3)_L\otimes
U(1)_X$ without particles with exotic electric charges is shown to be able to provide the quark mass
spectrum and their mixing, by means of universal see-saw mechanisms,
avoiding a hierarchy in the Yukawa coupling constants.}

\end{abstract}



The Standard Model (SM), with all its successes, is in the
unaesthetic position of having no explanation of fermion masses
and mixing angles, both in the quark and lepton sectors. Besides,
recent experimental results on neutrino oscillations \cite{altar},
which imply physics beyond the SM, call for extensions of the
model. In this regard, models based on the local gauge group
$SU(3)_c\otimes SU(3)_L\otimes U(1)_X$ (named in the literature
3-3-1 models) have been advocated recently, due to the fact that
several versions of the model can be constructed so that anomaly
cancellation is achieved \cite{pf,vl} under the condition that,
the number of families $N_f$ equals the number of colors $N_c=3$.
Among those models we have chosen to work with a particular one
that avoids the inclusion of fermion fields with exotic electric
charges.

As it has been recently pointed out \cite{peccei}, the appearance
of see-saw mechanisms could be in itself a guiding principle to
distinguish between fundamental scales and those which are not; if
this is so, then the explanation of the five orders of magnitude
spanned by the quark mass spectrum would require a new mass scale
unconnected with the electroweak symmetry breaking mass scale,
which may come from new physics, e.g.: supersymmetry, left-right
symmetric models, or something else. See-saw origin for all
fermion masses has been analyzed in the past in the context of
several models \cite{davidson}.

In the framework provided by a 3-3-1 model, and by using a
convenient set of Higgs fields, we show that one can avoid
hierarchies in the Yukawa couplings. The presence of a new scale
$V>>v$ related to the breaking of $SU(3)_L\otimes U(1)_X$,
triggers see-saw mechanisms that provide a sensible mass spectrum
for quarks. At the same time, these mechanisms provide relations
between the mass eingenstates and the weak interaction
eingenstates, and thus a Cabbibo Kobayashi Maskawa (CKM) mixing
matrix emerges.

The model based on the local gauge group $SU(3)_c\otimes
SU(3)_L\otimes U(1)_X$ has 17 gauge Bosons: one gauge field
$B^\mu$ associated with $U(1)_X$, the 8 gluon fields $G^\mu$
associated with $SU(3)_c$ which remain massless after breaking the
symmetry, and another 8 gauge fields associated with $SU(3)_L$ and
that we write for convenience as \cite{pgs}

\[{1\over 2}\lambda_\alpha A^\mu_\alpha={1\over \sqrt{2}}\left(
\begin{array}{ccc}D^\mu_1 & W^{+\mu} & K^{+\mu} \\ W^{-\mu} & D^\mu_2 &
K^{0\mu} \\ K^{-\mu} & \bar{K}^{0\mu} & D^\mu_3 \end{array}\right), \]
where $D^\mu_1=A_3^\mu/\sqrt{2}+A_8^\mu/\sqrt{6},\;
D^\mu_2=-A_3^\mu/\sqrt{2}+A_8^\mu/\sqrt{6}$, and
$D^\mu_3=-2A_8^\mu/\sqrt{6}$. $\lambda_i, \; i=1,2,...,8$, are the eight
Gell-Mann matrices normalized as $Tr(\lambda_i\lambda_j)
=2\delta_{ij}$.

The charge operator associated with the unbroken gauge symmetry
$U(1)_Q$ is given by $Q=\lambda_{3L}/2+
\lambda_{8L}/(2\sqrt{3})+XI_3$ where $I_3=Diag.(1,1,1)$ (the
diagonal $3\times 3$ unit matrix), and the $X$ values, related to
the $U(1)_X$ hypercharge, are fixed by anomaly cancellation. The
sine square of the electroweak mixing angle is given by
$S_W^2=3g_1^2/(3g_3^2+4g_1^2)$ where $g_1$ and $g_3$ are the
coupling constants of $U(1)_X$ and $SU(3)_L$ respectively, and the
photon field is given by
\begin{equation}\label{foton}
A_0^\mu=S_WA_3^\mu+C_W\left[\frac{T_W}{\sqrt{3}}A_8^\mu +
\sqrt{(1-T_W^2/3)}B^\mu\right],
\end{equation}
where $C_W$ and $T_W$ are the cosine and tangent of the electroweak mixing
angle, respectively.

The two weak flavor diagonal neutral currents in the model are coupled to the gauge Bosons 
\begin{equation}\nonumber \label{zetas}
Z_0^\mu=C_WA_3^\mu-S_W\left[\frac{T_W}{\sqrt{3}}A_8^\mu +
\sqrt{(1-T_W^2/3)}B^\mu\right]; \hspace{.1cm}
Z_0^{\prime\mu}=-\sqrt{(1-T_W^2/3)}A_8^\mu+\frac{T_W}{\sqrt{3}}B^\mu,
\end{equation}
where $Z_0^\mu$ coincides with the neutral gauge boson of the SM \cite{vl}. There is also an electrically neutral current associated with the flavor non diagonal gauge
boson $K^{0\mu}$ which is charged in the sense that it has a kind
of weak V isospin charge.

The quark content of the model is \cite{vl, pgs}:
$Q^i_{L}=(u^i,d^i,D^i)_L\sim(3,3,0),\;i=1,2$ for two families,
where $D^i_L$ are two extra quarks of electric charge $-1/3$ 
(numbers inside the parenthesis stand for the
$[SU(3)_c,SU(3)_L,U(1)_X]$ quantum numbers);
$Q^3_{L}=(d^3,u^3,U)_L\sim (3,3^*,1/3)$, where $U_L$ is an extra
quark of electric charge 2/3. The right handed quarks are
$u^{ac}_{L}\sim (3^*,1,-2/3),\; d^{ac}_{L}\sim (3^*,1,1/3)$ with
$a=1,2,3,$ a family index, $D^{ic}_{L}\sim (3^*,1,1/3),\;i=1,2$,
and $U^c_L\sim (3^*,1,-2/3)$.
The lepton content of the model is: $L_{aL} = (e_a^-,\nu_a^0,N_a^0)_L\sim (1,3^*,-1/3)$,
for $a=1,2,3=e,\mu,\tau$ respectively [three $SU(3)_L$ anti-triplets], and the three singlets
$e^+_{aL}\sim(1,1,1)$, with $\nu_a^0$ the neutrino field
associated with the lepton $e_a$ and $N_a^0$ playing the role of the corresponding 
right-handed neutrinos. There are not exotic charged leptons, and
universality for the known leptons in the three families is
present at tree level in the weak basis. With the former quantum
numbers the model is free of all the gauge anomalies\cite{pgs}.

Instead of using the set of Higgs fields introduced in the original papers \cite{vl}, we use the following set of four scalar triplets, with their Vacuum Expectation Values (VEV) as stated:
\begin{eqnarray*}\label{higgsses} 
\langle\phi_1^T\rangle &=&\langle(\phi^+_1, \phi^0_1,\phi^{'0}_1)\rangle =
\langle(0,0,v_1)\rangle \sim (1,3,1/3); \\  
\langle\phi_2^T\rangle &=&\langle(\phi^+_2, \phi^0_2,\phi^{'0}_2)\rangle =
\langle(0,v_2,0)\rangle \sim (1,3,1/3); \\ 
\langle\phi_3^T\rangle &=&\langle(\phi^0_3, \phi^-_3,\phi^{'-}_3)\rangle =
\langle(v_3,0,0)\rangle \sim (1,3,-2/3); \\ 
\langle\phi_4^T\rangle &=&\langle(\phi^+_4, \phi^0_4,\phi^{'0}_4)\rangle =
\langle(0,0,V)\rangle \sim (1,3,1/3), \\ 
\end{eqnarray*}
with the hierarchy $v_1\sim v_2\sim v_3\sim v\sim 10^2$ GeV $<< V\sim$ TeV.
The analysis shows that this set of VEV breaks the
$SU(3)_c\otimes SU(3)_L\otimes U(1)_X$ symmetry in two steps
following the scheme
\[SU(3)_c\otimes SU(3)_L\otimes U(1)_X\stackrel{V}{\longrightarrow}SU(3)_c\otimes SU(2)_L\otimes
U(1)_Y\stackrel{v}{\longrightarrow} SU(3)_c\otimes
U(1)_Q,\] 
where the first scale comes from $V+v_1\approx V$ and the second one from $v_2+v_3\approx v$. The breaking allows for the matching conditions: $g_2=g_3$ and
$1/g^{2}_Y=1/g_1^2+1/(3g_2^2)$, where $g_2$ and $g_Y$
are the gauge coupling constants of the $SU(2)_L$ and $U(1)_Y$
gauge groups in the SM.

Related models to this, with the same fermion content but different scalar sector ($\phi_1$ is absent) are analyzed in the papers in Refs.~\cite{vl}. Other 3-3-1 models without exotic electric charges, but with different fermion contents, can be found in Refs.~\cite{ozer}

The Higgs scalars introduced above are used to write the Yukawa
terms for the quarks. In the case of the Up quark sector, the most
general invariant Yukawa Lagrangian is given by
\begin{equation}\label{mup}
{\cal L}^u_Y=
\sum_{\alpha=1,2,4}Q_L^3\phi_\alpha C(h^U_\alpha
U_L^c+\sum_{a=1}^3h_{a\alpha}^uu_L^{ac})
+ \sum_{i=1}^2Q^i_L\phi_3^*
C(\sum_{a=1}^3h^u_{ia}u_L^{ac}+h_i^{\prime U}U_L^c)
+ h.c.,
\end{equation}
where the $h^{u,U\prime}$s are couplings that we assume of order
one. $C$ is the charge conjugation
operator. In order to restrict the number of Yukawa couplings, and
produce a realistic fermion mass spectrum, we introduce the following anomaly-free \cite{ross} discrete $Z_2$ symmetry
\begin{equation}\label{z2}
Z_2(Q^a_L,\phi_2,\phi_3,\phi_4,u^{ic}_L,d^{ac}_L)=1 \,\,\,;\,\,\,
Z_2(\phi_1, u^{3c}_L,U^c_L,D^{ic}_L, L_{aL}, e^+_{aL})=0,
\end{equation}
where $a=1,2,3 (=e,\mu,\tau$ for the leptons) and $i=1,2$ are family indices.

Then in the basis $(u^1,u^2,u^3,U)$ we get, from Eq.(\ref{mup}-\ref{z2}),
the following tree-level Up quark mass matrix:

\begin{equation} \label{matrixU}
M_u=\left(\begin{array}{cccc}
0 & 0 & 0 & h^u_{11}v_1\\
0 & 0 & 0 & h^u_{21}v_1\\
h^u_{13}v_3 & h^u_{23}v_3 & h^u_{32}v_2 & h_{34}^uV\\
h_1^{\prime U}v_3 & h^{\prime U}_2v_3 & h^U_2v_2 & h^U_4V \\
\end{array}\right), \end{equation}
which is a see-saw type mass matrix, with one eigenvalue equal to
zero.

On the other hand, the Yukawa terms for the Down quark sector,
using the four Higgs scalars introduced in Eq. (\ref{higgsses}),
are:

\begin{equation}\label{mdown}
{\cal L}^d_Y = \sum_{\alpha=1,2,4}
\sum_{i}Q^i_L\phi_\alpha^*C(\sum_ah^d_{ia\alpha}d_L^{ac}
+\sum_jh^D_{ij\alpha}D_L^{jc})
+ Q_L^3\phi_3C(\sum_ih^D_iD_L^{ic}+\sum_{a}h_a^dd_L^{ac})+h.c..
\end{equation}

In the basis $(d^1,d^2,d^3,D^2,D^3)$ and using the discrete
symmetry $Z_2$, the former expression produces the following
tree-level Down quark mass matrix:
\begin{equation}\label{MD} M_d=\left(\begin{array}{ccccc}
0 & 0 & 0 & h_{11}^{d}v_1 & h_{21}^{d}v_1 \\
0 & 0 & 0 & h_{12}^{d}v_1 & h_{22}^{d}v_1 \\
0 & 0 & 0 & h_{13}^{d}v_1 & h_{23}^{d}v_1 \\
h_{11}^{D}v_2 & h_{21}^{D}v_2 & h_{1}^{D}v_3 & h_{114}^DV & h_{214}^DV \\
h_{12}^{D}v_2 & h_{22}^{D}v_2 & h_{2}^{D}v_3 & h_{124}^DV & h_{224}^DV \\
\end{array}\right),
\end{equation}
where we have used
$h^{D(d)}_{ia\alpha}v_\alpha=h^{D(d)}_{ia}v_\alpha$. The mass matrix $M_d$ is
again a see-saw type, with at least one eigenvalue equal to zero.

Before entering into a more detailed analysis of $M_u$ and $M_d$,
let us insist in the resulting see-saw character of these
matrices. In both cases there is a zero eigenvalue that we
immediately identify with the $u$ and $d$ quarks of the first
family, respectively. Then, they are massless at tree-level in the
model considered here. In the U sector, the $c$ quark acquires a
see-saw mass, while in the D sector, both $s$ quark and $b$ quark
have see-saw masses (nevertheless, with a particular election of
parameters, one can end up with a massless $s$ quark too). The U
sector structure is in some sense singular because the top mass is
of the order of the electroweak scale; in fact it gets already a
tree level mass of this order.

A further numerical check of the matrices is definitive
in the sense that the model provides a see-saw mass hierarchy
defined by the relationship between the symmetry breaking scales
$v/V$.  
In what follows, and without loss of generality, we are going to
impose the condition $v_1=v_2=v_3\equiv v << V$, with the value
for $v$ fixed by the mass of the charged weak gauge boson
$M_{W^\pm}^2=g_3^2(v_2^2+v_3^2)/2=g_2^2v^2$ which implies
$v=246/2=123$ GeV ($v_1$ is associated with an $SU(2)_L$ singlet 
and does not contribute to the $W^\pm$ mass).

Starting with the $U$ matrix, the analysis shows that
$M_u^\dagger M_u$ has one zero eigenvalue, related to the
eigenvector $[(h_{32}^uh_{2}^{\prime U}-h_{23}^uh_{2}^U),
(h_{13}^uh_{2}^U-h_{32}^uh_{1}^{\prime U}), (h_{23}^uh_{1}^{\prime
U}-h_{13}^uh_{2}^{\prime U}),0]$, that we may identify with the up
quark $u$ in the first family, which remains massless at
tree-level.

In order to simplify the otherwise cumbersome calculations and to
avoid the proliferation of unnecessary parameters at this stage of
the analysis, we propose to start with the following simple matrix

\begin{equation} \label{SU}
M^\prime_u=hv\left(\begin{array}{cccc}
0 & 0 & 0 & 1\\
0 & 0 & 0 & 1\\
1 & 1 & h_{32}^u/h & \delta^{-1} \\
1 & 1 &  1  & \delta^{-1} \\
\end{array}\right),
\end{equation}
where $\delta=v/V$ is the expansion parameter and $h$ is a
parameter that can take any value of order $1$. The results below
show that this matrix has the necessary ingredients to produce a
consistent mass spectrum.

Neglecting terms of order $\delta^5$ and higher, the four
eigenvalues of $M_u^{\prime\dagger} M^\prime_u$ are: one zero
eigenvalue related to the eigenstate $(u^1-u^2)/\sqrt{2}$ 
(notice the maximal mixing present); a see-saw eigenvalue $
4h^2V^2\delta^4=4h^2v^2\delta^2\approx m_c^2$ associated to the
charm quark, and the other two
\[ \frac{h^2V^2\delta^2}{2}[e_-^2+\delta^2e_+^2(4-e_-^2)/4]
\approx \frac{v^2}{2}(h-h_{32}^u)^2\approx m_t^2\]
\[h^2V^2[2+\delta^2(6+e_+/2)+\delta^4(4e_+^2-e_+^2e_-^2-32)/8] \approx m_U^2\]
where $e_{\pm}=(1 \pm h_{32}^u/h)$. The eigenvectors are given by the
rows of the following $4\times 4$ unitary matrix:
\begin{equation}\label{mixU}
U^U_L = \left(\begin{array}{cccc}
\frac{1}{\sqrt{2}} &-\frac{1}{\sqrt{2}} & 0 & 0\\
\frac{C_{\eta_1}}{\sqrt{2}} &\frac{C_{\eta_1}}{\sqrt{2}}  & 0 & -\frac{S_{\eta_1}}{\Delta}\\
0 & 0 & \Delta^{-1} & -\frac{\delta\, e_+}{2\Delta} \\
\frac{S_{\eta_1}}{\sqrt{2}} & \frac{S_{\eta_1}}{\sqrt{2}} &
\frac{C_{\eta_1} \delta\, e_+}{2\Delta}
& \frac{C_{\eta_1}}{\Delta} \\
\end{array}\right),
\end{equation}
where $C_{\eta_1}$ and $S_{\eta_1} \approx
\sqrt{2}\delta(1-3\delta^2)$ are the cosine and sine of a mixing
angle $\eta_1$, and $\Delta = \sqrt{(1+\delta^2e_+/4)}$.

So, in the Up quark sector the heavy quark gets a large mass of
order $V$, the top quark gets a mass at the
electroweak scale (times a difference of Yukawas that in the
general case of matrix (\ref{matrixU}) is $(h_2^U-h_{32}^u)$),
the charm quark gets a see-saw mass, and the first family up quark
$u$ remains massless at tree-level. From the former expressions we get $|h_2^U-h_{32}^u|\sim 2$ and $m_c\approx 2hv^2/V$, which in turn implies $V\approx hm_t^2/m_c\approx 19.4 h$ TeV., fixing in this way an upper limit for the 3-3-1 mass scale $V$ 
(experimental values are taken from Ref. \cite{pdg}).

We go now to the D quark mass matrix. This matrix is full of
physical possibilities, depending upon the particular values
assigned to the Yukawa couplings. For example, if all of them are
different from each other, then the matrix $M_d^\dagger M_d$ has
rank one with a zero eigenvalue related to the eingenvector
$[(h_{22}^D h_{1}^D-h_{21}^D h_{2}^D), (h_{11}^D h_{2}^D -
h_{12}^D h_{1}^D), (h_{21}^D h_{12}^D -h_{11}^D h_{22}^D),0,0]$,
that we may identify with the down quark $d$ in the first family
(which in any case remains massless at tree-level); for this case
the general analysis shows that we have two see-saw eigenvalues
associated with the bottom $b$ and strange $s$ quarks.

In the particular case when all the Yukawas are equal to one but
$h_{114}^D=h_{224}^D=H^D\neq 1$, the null space of $M_d^\dagger
M_d$ has rank two, with the eigenvectors associated with the zero
eigenvalues given by $[-2,1,1,0,0]/\sqrt{6}$ and
$[0,-1,1,0,0]/\sqrt{2}$, which in turn implies only one see-saw
eigenvalue associated with the bottom quark $b$, with a value for
its mass approximately equal to $6v\delta/(1+H^D)$, with masses
for the two heavy states of the order of $V(1\pm H^D)$.

For the first case mentioned, the chiral symmetry remaining at
tree-level is $SU(2)_f$ (quarks $u$ and $d$ massless), and for
the second case the chiral symmetry is $SU(3)_f$ (quarks $u$, $d$
and $s$ are massless). In both cases the chiral symmetry is broken by the
radiative corrections.

In any way, a realistic analysis of the down sector requires to
have in mind the mixing matrix (\ref{mixU}) of the up quark sector
and the fact that the CKM mixing matrix is almost unitary and diagonal. Aiming
at this and in order to avoid again a proliferation of parameters,
let us analyze the particular case given by the following
left-right symmetric (hermitian) down quark mass matrix:

\begin{equation} \label{matrixD}
M^\prime_d=h^\prime v\left(\begin{array}{ccccc}
0 & 0 & 0 & 1 & 1 \\
0 & 0 & 0 & 1 & 1 \\
0 & 0 & 0 & f & g \\
1 & 1 & f & H^D\delta^{-1} & \delta^{-1} \\
1 & 1 & g & \delta^{-1} & H^D\delta^{-1} \\
\end{array}\right),
\end{equation}
where $f$ and $g$ are parameters of order one. This is the most
general hermitian mass matrix with only one eigenvalue equal to
zero, related with the state $(d^1-d^2)/\sqrt{2}$, as required in
order to end up with an almost diagonal CKM mixing matrix.

The two see-saw exact eigenvalues of $M^\prime_d$ are:
\begin{equation}
- h^\prime\,v\, \frac{\delta}{4}
\left\{\left[\frac{(f-g)^2}{H^D - 1} + \frac{8
+ (f+g)^2}{1+ H^D}\right]
\pm\sqrt{\left[\frac{(f-g)^2}{H^D - 1} + \frac{8 +
(f+g)^2}{1+ H^D}\right]^2 - \frac{8 (f-g)^2}{1-(H^D)^2}}\right\}.
\end{equation}

Moreover, notice that for the particular case $g = -f$, the five
eigenvalues of the hermitian matrix above get the following simple
exact analytical expressions

\begin{equation}
 \frac{h^{\prime}\,\delta^{-1}\,v}{2}
\left[0, H^D_+(1\pm \sqrt{1 + 16 \delta^2/(H^D_+)^2}),\;
H^D_-(1\pm \sqrt{1 + 8 f^2
\delta^2/(H^D_-)^2})\right] \nonumber,
\end{equation}
where $H^D_{\pm} = H^D \pm 1$. The see-saw values are thus
$-4\,\delta h^\prime v/H^D_+$ and $- 2\, \delta\, f^2\, h^\prime \, v/H^D_-$; they imply $f^2h^\prime/h\approx m_b H_-^D/m_c\approx 3H_-^D$ and $2h^\prime/h\approx H^D_+ m_s/m_c$, that can be seen as either a mild hierarchy between $h$ and $h^\prime$, or implying a detailed tuning of some of the parameters of order one (inconvenience that could be avoided by working in a frame where $SU(3)_f$ becomes the original chiral symmetry).

The eigenvectors are now given by the rows of the following $5 \times 5$ unitary matrix
\begin{equation}\label{mixD}
U^D_L = \frac{1}{\sqrt{2}}\,\left(\begin{array}{ccccc} 1 & -1 & 0
& 0 & 0 \\
C_{\eta_2} & C_{\eta_2} & 0 &
-S_{\eta_2} & S_{\eta_2} \\
0 & 0 & \sqrt{2}\,C_{\eta_3} & - S_{\eta_3}  &  - S_{\eta_3}\\
S_{\eta_2} &  S_{\eta_2} & 0 & C_{\eta_2} & -C_{\eta_2} \\
0 & 0 &  \sqrt{2}\,S_{\eta_3} & C_{\eta_3} & C_{\eta_3}\\
\end{array}\right),
\end{equation}
where: $C_{\eta_2}$ and $S_{\eta_2} \approx 2\,\delta [1 -
\delta^2/(H^D_+)^2]/H^D_+ $; $C_{\eta_3}$ and $S_{\eta_3} \approx
\sqrt{2}\,\delta\,f[1 - 3\,f^2 \delta^2/(H^D_-)^2]/H^D_- $ are the
cosines and sines of other two mixing angles $\eta_2$ and $\eta_3$.
Notice that up to this point, the CKM matrix $U_{CKM}^{(0)} =
U^{u\,\dagger}_L\,U^d_L$ deviates from the identity just by terms
of the order $\delta^2$ and higher; where $U^u_L$ is the $3 \times
3$ upper sector of $U^U_L$ of eq.(\ref{mixU}) and the same for
$U^d_L$.

\begin{figure}
\onefigure{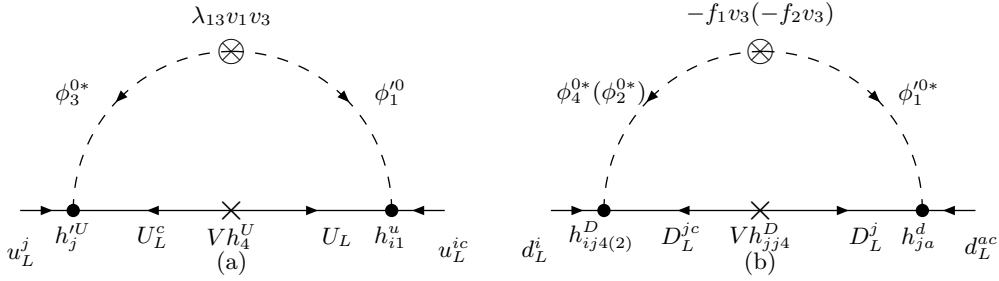}
\caption{The five one-loop diagrams that produce the radiative masses for the quarks $u$ and $d$.}
\label{fig1}
\end{figure}

The consistency of the model requires that one can identify
mechanisms able to produce masses for the first family, and to
generate the CKM mixing angles. A detailed study of the Lagrangian
for the Up quark sector (\ref{mup}), the discrete $Z_2$ symmetry (\ref{z2})
and the scalar potential, allows us to draw the radiative diagram in Fig.(\ref{fig1}a),
which is the only diagram available to produce a finite one-loop radiative
correction in the quark subspace $(u^1,u^2)$ of the Up quark
sector. The mixing of the Higgs Bosons comes from a term in the
scalar potential of the form
$\lambda_{13}(\phi^*_1\phi_1)(\phi^*_3\phi_3)$, 
which turns on the radiative correction.

In order to have a contribution different from zero we must avoid
maximal mixing in the first two weak interaction states, otherwise
a submatrix of the democratic type arises. This is done by
taking $h^u_{11} = 1-k$ and $h^{\prime U}_1 = 1 +k$ in matrix (\ref{SU}) insted of 1, where $k$ must be a very small parameter inorder to guarantee the see-saw character of the Up sector quark mass matrix.
Evaluate the contribution coming from the diagram in Fig. (\ref{fig1}a) we get

\begin{equation}\label{diag1} \nonumber
\Delta_{ji}=N_{ji}[M^2m_1^2\ln (M^2/m_1^2) - M^2m_3^2\ln (M^2/m_3^2)
+m^2_3m_1^2\ln (m^2_1/m_3^2)],
\end{equation}
where $N_{ji}=h_j^{\prime
U}h_{i1}^u\lambda_{13}v_1v_3M/[16\pi^2(m_3^2-m_1^2)
(M^2-m_1^2)(M^2-m_3^2)],\; M=h_4^UV$ is the mass of the heavy Up
quark, and $m_1$ and $m_3$ are the masses of $\phi_1^{\prime 0}$
and $\phi_3^0$ respectively. To estimate the contribution given by
this diagram we assume the validity of the ``extended survival
hypothesis" \cite{esh} which in our case means $m_1\approx
m_3\approx v$, implying
\begin{equation}
m_u \approx \,\lambda_{13}v\delta\ln (V/v)/8\pi^2\approx
0.85 \lambda_{13} \, \mbox{MeV},
\end{equation} 
which for $\lambda_{13} \sim  2 $ produces $ m_u\approx 1.7 $ MeV, which
is of the correct order of magnitude~\cite{pdg} (result independent of the value of $k$ in first approximation).
Due to the fact that the parameter $k \neq 0$, the
state related to the $u$ quark looses its maximal mixing, becoming
now $ \{-(h - h_{32}^u) u^1 + [h- h_{32}^u(1 -k)] u^2 + k u^3\}/N$, with $N$
being the normalization factor. The value of $k$ is estimated with
the value of the Cabbibo angle to be $k \approx 0.1$.

For the Down quark sector there are four one-loop diagrams, two for $D^1$ and other two for $D^2$ as depicted in Fig.(\ref{fig1}b). The mixing in the Higgs sector comes from terms in the scalar potential of the form $f_1\phi_1\phi_3\phi_4+f_2 \phi_1\phi_2\phi_3+h.c.$. When the algebra is done we get
\begin{equation}
m_d\approx 2(f_1+f_2)\delta\ln (V/v)/8\pi^2,
\end{equation}
which for $f_1=f_2\approx v$ implies $m_d\approx 2m_u$, without introducing a new mass scale in the model.

The discrete $Z_2$ symmetry introduced eliminates possible tree-level lepton mass terms of the form $L_{aL}\phi_3Ce_{bL}$ and $L_{aL}L_{bL}\phi_3$. Then in order to generate masses for the leptons we must use either leptoquark Higgs Fields if we intent to use the radiative mechanism, or exotic leptons if we want to use see-saw mechanisms. For the neutrinos for example this analysis has been done in Ref.\cite{kita}, where new $SU(3)_L$ Higgs scalar multiplets are introduced.

In a model like this with four scalar triplets, we should worry about possible flavor changing neutral current (FCNC) effects. First we notice that due to our $Z_2$ symmetry, they do not occur at tree-level because each flavor couples only to a single multiplet. They can enter as a consequence of the violation of unitarity of the CKM matrix $U_{CKM}^0$ which is a $3\times 3$ submatrix of a rectangular $4\times 5$ matrix. The violation of unitarity in our analysis is proportional to $\delta^2$, implying FCNC proportional to $\delta^4$. Then, a value of  $\delta \approx 10^{-2}$ is perfectly safe as far as violation of unitarity of the CKM matrix and possible FCNC effects are concerned. Experimental constraints on the posible violation of unitarity of the CKM matrix are discussed in Section 11 of Ref.~\cite{pdg}.

In several 3-3-1 models with three scalar triplets \cite{pf, vl} a discrete symmetry can suppress mass terms for the neutral Higgs bosons and to produce axion states \cite{ple}. The preliminary analysis shows that the $Z_2$ symmetry introduced in our model with four scalar triplets, provides only with the eight Goldstone Bosons needed, and nothing else.

In conclusion, we have presented a model with only two energy scales, that has the power of avoiding
hierarchies among Yukawa couplings. Throughout the analysis, all
the Yukawas are of order one, as also is the case for the dimensionless Higgs
coupling $\lambda_{13}$. The new ingredients of the model are: the
mass scale $V$ used to define the expansion parameter $\delta$, a
new set of Higgs scalars and VEV and the discrete anomaly-free symmetry $Z_2$. All
this triggers generalized see-saw mechanism in the Up and Down quark sectors.

\end{document}